\documentclass[10pt,prc,showpacs,superscriptaddress,twocolumn]{revtex4}
\usepackage{graphicx}
\begin{document}
\title{Elliptic Flow Analysis at RHIC: Fluctuations vs. Non-Flow Effects}
\author{Xianglei Zhu}
\affiliation{Frankfurt Institute for Advanced Studies (FIAS), Max-von-Laue-Str.~1, D-60438 Frankfurt am Main,
Germany} \affiliation{Institut f\"ur Theoretische Physik, Johann Wolfgang Goethe-Universit\"at,
Max-von-Laue-Str.~1, D-60438 Frankfurt am Main, Germany} \affiliation{Physics Department, Tsinghua University,
Beijing 100084, China}
\author{Marcus Bleicher}
\affiliation{Institut f\"ur Theoretische Physik, Johann Wolfgang Goethe-Universit\"at, Max-von-Laue-Str.~1,
D-60438 Frankfurt am Main, Germany}
\author{Horst St\"ocker}
\affiliation{Frankfurt Institute for Advanced Studies (FIAS), Max-von-Laue-Str.~1, D-60438 Frankfurt am Main,
Germany} \affiliation{Institut f\"ur Theoretische Physik, Johann Wolfgang Goethe-Universit\"at,
Max-von-Laue-Str.~1, D-60438 Frankfurt am Main, Germany}
\begin{abstract}
The cumulant method is applied to study elliptic flow ($v_2$) in Au+Au collisions
at $\sqrt{s}=200$AGeV, with the UrQMD model. In this approach, the true event plane is known and both the
non-flow effects and event-by-event spatial ($\epsilon$) and $v_2$ fluctuations exist. Qualitatively, the hierarchy
of $v_2$'s from two, four and six-particle cumulants is consistent with the STAR data, however, the magnitude of
$v_2$ in the UrQMD model is only 60\% of the data. We find that the four and six-particle cumulants are good
measures of the real elliptic flow over a wide range of centralities except for the most central and very
peripheral events. There the cumulant method is affected by the $v_2$ fluctuations. In mid-central collisions,
the four and six-particle cumulants are shown to give a good estimation of the true differential $v_2$,
especially at large transverse momentum, where the two-particle cumulant method is heavily affected by the
non-flow effects. 
\end{abstract}
\maketitle

To create extremely hot and dense matter with partons as its
fundamental components - called the Quark-Gluon Plasma (QGP) - is a
major goal of  current and future high energy heavy-ion collisions
experiments at SPS, RHIC and LHC \cite{QM2005}. However, due to the
complex nature of the relativistic nucleus-nucleus reactions, the
QGP, if it has been created, escapes direct detection. Therefore, in
order to distinguish the existence and later on to investigate the
properties of the new state of matter, one must find observables
which allow to deduce the properties of the intermediate (QGP)
state from the final state hadrons. Elliptic flow ($v_2$), which is
the second Fourier harmonic \cite{Voloshin} in the transverse distribution of the
emitted particles, is expected to be sensitive to the early pressure
gradients and therefore the equation of state (EOS) of the formed
fire-ball in the heavy-ion collisions \cite{Voloshin,Ollitrault}. Recent
elliptic flow results on  Au+Au collisions at $\sqrt{s}=200$~AGeV
\cite{STAR_flow,PHENIX_PRL94,PHENIX_V2,PHOBOS_V2} indeed indicate high pressure
gradients in the early stage of the reaction and might therefore
hint towards the existence of an intermediate QGP state at this
energy.

In principle, the elliptic flow of hadrons at low transverse momenta
($p_T$) can be related to the degree of thermalization, the
viscosity  and the EOS of the produced matter
\cite{Ollitrault,Teaney_Hydro,hydro}. On the other hand, the
elliptic flow of the high $p_T$ particles is related to jet
fragmentation and  energy loss of the primordially produced hard
Anti-quark-Quark pair when traveling through the hot QCD medium
\cite{Gyulassy_Jet}. At low $p_T$, results from most of the RHIC
experiments \cite{STAR_flow,PHENIX_PRL94,PHENIX_V2,PHOBOS_V2} indicate a gradual
increase of $v_2$ with the increase of $p_T$. This behavior  is
approximately consistent with the prediction of relativistic
hydrodynamical calculation with a first order phase transition to a
QGP \cite{hydro}. When $p_T\ge 1.5$~GeV/c, the $v_2$ begins to saturate and
eventually decreases \cite{STAR_flow,STAR_highpt}. This is a clear
signal for the breakdown of hydrodynamics at intermediate $p_T$ and
a transition towards jet physics. As shown in
\cite{Gyulassy_Jet}, the $v_2$ at large $p_T$ might be a
sensitive probe of the initial parton density distribution of the
Quark-Gluon matter produced. Thus an accurate $v_2$ measurement
might allow deeper insights into the bulk properties of the produced
matter.

However, an unambiguous experimental measurement of the elliptic
flow is not a trivial task due to the unknown orientation of the
reaction plane. Often, experiments use the so called reaction plane
method \cite{Poskanzer_Voloshin} to extract the magnitude of the
elliptic flow. In this method, the reaction plane is fixed according
to the flow vector of the event, then the estimated $v_2$ with
respect to the chosen reaction plane is corrected for the event
plane resolution, which accounts for the error in the deduction of
the reaction plane. The original reaction plane method is consistent
with the two-particle correlation method
\cite{Poskanzer_Voloshin,STAR_flow,STAR_PRC66}, in which $v_2$ is
related to the two particle angular difference $\phi_1-\phi_2$ by
$v_2=\sqrt{\langle cos2(\phi_1-\phi_2)\rangle}$. However, these
two-particle correlations based methods might suffer from effects which are not
related to the reaction plane, these additional contributions are
usually called non-flow effects \cite{Borghini:2000cm}, such as the overall transverse momentum conservation, small angle azimuthal correlations due to final state interactions, resonance decays, jet production \cite{Kovchegov:2002nf} and quantum correlations due to the HBT effect \cite{Dinh:1999mn}. In order to eliminate the non-flow contributions to the measured collective flow in the reaction plane method, a rapidity gap between the particles used to estimate the reaction plane and the measured particles is usually introduced. But whether this improvement works well is still not clear. Recently, the
cumulant method was proposed \cite{Borghini} to diminish the non-flow effects. The idea of the
cumulant method is to extract flow with many-particle cumulants, which are the many-particle correlations with subtraction of the contributions from the correlations due to the lower-order multiplets. It is believed that the pure many-particle non-flow correlations have much less contributions to the measured flow in the many-particle cumulant method.
In other words, the many-particle cumulant method should be much less sensitive to
non-flow effects \cite{Borghini}. At RHIC energy, the cumulant method has been
applied by STAR \cite{STAR_flow,STAR_PRC66,STAR_Voloshin} and PHENIX \cite{PHENIX_PRL94} in the flow analysis of Au+Au collisions. It is found that the integral $v_2$
from the two-particle correlations which is denoted as $v_2\{2\}$ is
about 15\% larger than the values from four and six-particle
cumulants($v_2\{4\}$ and $v_2\{6\}$). One might attribute the extra $v_2$ in $v_2\{2\}$ to the non-flow correlations and conclude that the
non-flow contribution has been successfully eliminated in the
results with four or six-particle cumulants \cite{STAR_flow}.
However, as indicated in \cite{miller03,STAR_PRC66}, the $v_2$
from many-particle cumulants is also affected by the event-by-event
$v_2$ fluctuations. For a rough estimation of the fluctuations'
contribution to the measured elliptic flow, the reader is referred
to \cite{miller03}. There, the estimation is based on the assumption
that the $v_2$ of an event is proportional to initial eccentricity
of the nucleons or quarks \cite{Sorge_Heiselberg_Voloshin}. The authors found that the difference
between $v_2\{2\}$ and $v_2\{4\}$ can also be explained by a
definite amount of the fluctuations of $v_2$ which gives a larger
$v_2\{2\}$ and a smaller $v_2\{4\}$ than the exact $v_2$ for the semi-central collisions. Here it is worth stressing that this conclusion can be extended to the cumulant analysis of any order harmonics $v_n$ with $n\ge 1$, i.e. the $v_n$ fluctuations result in $|v_n\{2\}|> |v_n\{4\}|$ generally. Therefore, both non-flow correlations and $v_2$ fluctuations can explain the fact that $v_2\{2\}>v_2\{4\}$ at RHIC. If the non-flow effects are dominant, the exact $v_2$ is $v_2\{4\}$ or $v_2\{6\}$; if the fluctuations are dominant, the exact $v_2$ is $(v_2\{2\}+v_2\{4\})/2$ according to \cite{miller03}. This ambiguity poses large systematic error to the $v_2$ measurement especially at large $p_t$, where the difference between $v_2\{2\}$ and $v_2\{4\}$ is most prominent. So it is essential to make clear which effect is dominant in the difference between $v_2\{2\}$ and $v_2\{4\}$ in order to reduce the large systematic uncertainty.

In this article, we use the UrQMD model (v2.2) \cite{urqmd,urqmd2.1} to test the robustness of the cumulant method for
the elliptic flow analysis. The advantages of using a transport approach compared to hydrodynamics are immanent:
\begin{itemize}
\item
Firstly, transport models do not make any additional assumptions on local/global equilibration of the matter
created during the collisions, but treat the non-equilibrium processes directly.
\item
Secondly, the present transport approach includes most non-flow correlations such as the overall transverse momentum conservation, small angle azimuthal correlations due to final state interactions, resonance decay and jet prodcution, naturally during the systems evolution. 
\item
Thirdly, the UrQMD model is an event by event model, hence it
contains the event by event fluctuations of the elliptic flow.
\end{itemize}
Finally, the reaction plane angle $\Phi_R$ is known in the model, which allows the direct calculation of the exact elliptic
flow from its basic definition, that is $v_2=\langle \cos2(\phi-\Phi_R)\rangle$. Therefore, the UrQMD model, even if for the time being still underpredicts the
integral $v_2$ in $\sqrt{s}$=200AGeV Au-Au collisions at RHIC, is an ideal tool to find out
whether the $v_2$ fluctuations and non-flow effects have large effects on the experimentally used cumulant
method.

\begin{figure}[h]
\centering
\includegraphics[totalheight=2.3in]{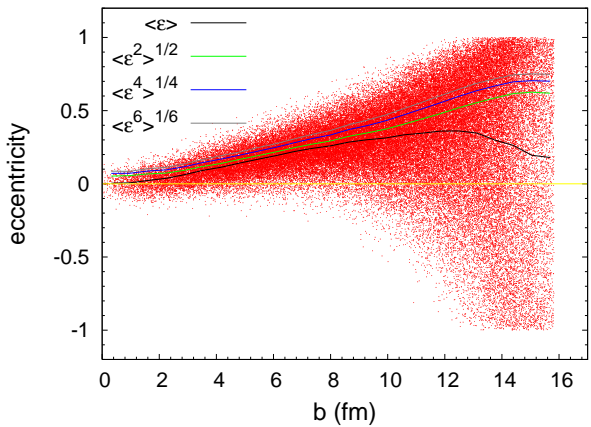}
\caption{(Color online) Scatter plot of the initial spatial eccentricities ($\epsilon=\frac{\langle y^2\rangle
-\langle x^2\rangle}{\langle y^2\rangle +\langle x^2\rangle}$) of the participants at different impact
parameters from the UrQMD model. The black, green, blue and grey lines are the average eccentricity
$\left\langle\epsilon\right\rangle$, $\left\langle\epsilon^2\right\rangle^{1/2}$,
$\left\langle\epsilon^4\right\rangle^{1/4}$ and $\left\langle\epsilon^6\right\rangle^{1/6}$ respectively.}
\label{figesca}
\end{figure}

\begin{figure}[h]
\centering
\includegraphics[totalheight=2.3in]{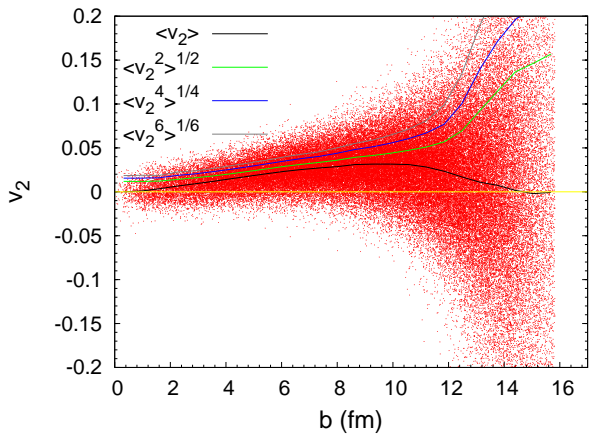}
\caption{(Color online) Scatter plot of the event $v_2$ averaged over all particles within $|\eta|<2.5$ at
different impact parameters from the UrQMD model. The black, green, blue and grey lines are the average elliptic
flow  $\left\langle v_2\right\rangle$, $\left\langle v_2^2\right\rangle^{1/2}$, $\left\langle
v_2^4\right\rangle^{1/4}$ and $\left\langle v_2^6\right\rangle^{1/6}$ respectively.} \label{figsca}
\end{figure}

\begin{figure}[h]
\centering
\includegraphics[totalheight=2.3in]{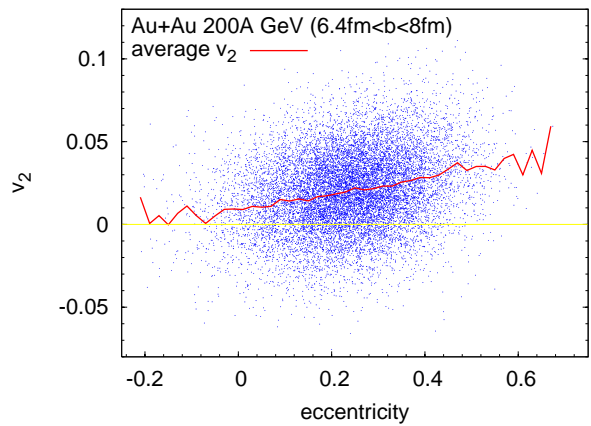}
\caption{(Color online) Scatter plot of the event $v_2$ averaged over all particles within $|\eta|<2.5$ at
different eccentricities from the UrQMD model. The red line is the average elliptic flow.} \label{v2s2}
\end{figure}

Before the application of the cumulant method, let us begin by examining the magnitude of the fluctuations of
eccentricity and $v_2$. Fig.\ref{figesca} shows a scatter plot of initial spatial eccentricity
($\epsilon=\frac{\langle y^2\rangle -\langle x^2\rangle}{\langle y^2\rangle +\langle x^2\rangle}$) of the
participants as a function of impact parameter based on a subset of the available UrQMD minimum-bias events. As
one can see, the eccentricity fluctuations in the model are in magnitude similar to the eccentricity itself. Note
that the magnitude of the fluctuations is quite similar to the estimates calculated with a Monte Carlo Glauber
model \cite{miller03}. Due to the large event by event fluctuations,
$\left\langle\epsilon^2\right\rangle^{1/2}$, $\left\langle\epsilon^4\right\rangle^{1/4}$ and
$\left\langle\epsilon^6\right\rangle^{1/6}$ are much larger than $\left\langle\epsilon\right\rangle$, especially
for the most central and peripheral events. The eccentricity fluctuations are supposed to be the main origin of
the $v_2$ fluctuations. Fig.\ref{figsca} is the scatter plot of the event $v_2$ averaged over all particles with
$|\eta|<2.5$, as a function of impact parameter based on the same subset of UrQMD minimum-bias events. Like the
eccentricity, the event $v_2$ fluctuations are also of the same magnitude as the $v_2$ itself. Therefore,
$\left\langle v_2^2\right\rangle^{1/2}$, $\left\langle v_2^4\right\rangle^{1/4}$ and $\left\langle
v_2^6\right\rangle^{1/6}$ are also much larger than $\left\langle v_2\right\rangle$, especially in the most
central and very peripheral centralities where the $\left\langle v_2\right\rangle$ is very small.
Fig.\ref{v2s2} shows the correlation of $v_2$ and eccentricity in the semi-central collisions. The fact that the $\left\langle v_2 \right\rangle$ is proportional to eccentricity, shows that the fluctuations in the initial conditions are the major origin of $v_2$ fluctuations \cite{Sorge_Heiselberg_Voloshin} besides the statistical noise \cite{Mrowczynski:2002bw} due to the limited number of particles used to estimate $v_2$. 

The observation of these large fluctuations puts some doubt on the accuracy of the experimental methods for the
extraction of the elliptic flow parameters. Therefore, we will now focus on the cumulant method and compare the
model results (with fluctuations and non-flow effects) obtained by different order cumulant methods with the exact $v_2$. For the detailed application of the cumulant method, the reader is referred to
\cite{Borghini}. In our analysis we use unit weights in the evaluation of the
generating function of the cumulants. The parameter $r_0$ is 1.5 as usually used in previous analysis. Actually,
the detailed investigation indicates that the present results are rather insensitive to the $r_0$ values as
pointed out before in \cite{Borghini}. For the integral $v_2$ analysis, we use all particles in the
pseudorapidity region $|\eta|<2.5$, and the number of particles from event to event fluctuates in each
centrality bin. We have tested the cumulant method with fixed number of particles in each event for the same
centrality and find that the results do not change within the present statistical error. The centralities in our
analysis are selected according to the same geometrical fractions of the total cross section
(0-5\%,5-10\%,10-20\%,20-30\%,30-40\%,40-50\%,50-60\%,60-70\%) as used by the STAR experiment \cite{STAR_flow},
however, we use impact parameter cuts instead of multiplicity cuts. More than $1.3\cdot 10^6$ minimum bias
events are used in the integral $v_2$ analysis. In order to increase the statistics at the most peripheral
centrality bin(60-70\%), additional $7 \cdot 10^5$  events are added in this centrality bin.

\begin{figure}[h]
\centering
\includegraphics[totalheight=2.3in]{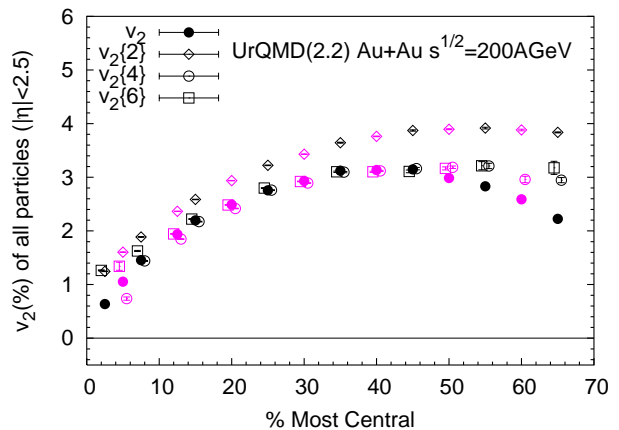}
\caption{(Color online) The integral $v_2$ results($v_2\{2\}$,$v_2\{4\}$ and $v_2\{6\}$) from the cumulant
method are compared to the exact $v_2$ in different centrality bins. The pink (grey) points are the
corresponding results from the enlarged centrality bins which merge two of the original bins.} \label{fig1}
\end{figure}

Fig.\ref{fig1} shows the calculated integral $v_2$ results as a function of centrality. 
The elliptic flow parameters extracted from two-particle cumulant $v_2\{2\}$ deviate
rather strongly from the theoretically expected $v_2$ as obtained from the known reaction plane. In fact, at all centralities, $v_2\{2\}$ is larger than the exact $v_2$ by 18\%. However, for mid-central
collisions ($\sigma/\sigma_{\rm tot}\sim 10-50\%$), the elliptic flow from four particle
($v_2\{4\}$) and six particle cumulants ($v_2\{6\}$) show almost no difference and both agree well with the
exact $v_2$. This fact is quite contrary to the prediction in \cite{miller03}, that is, the exact $v_2$ should be in the middle of the $v_2\{2\}$ and $v_2\{4\}$ if the differences between the cumulant methods are mainly due to $v_2$ fluctuations. Therefore, we conclude that for semi-central to semi-peripheral centralities the contribution of the $v_2$ fluctuations to the cumulant results is almost negligible and the difference between $v_2\{2\}$ and $v_2\{4\}$ or $v_2\{6\}$, is mainly due to non-flow effects in the UrQMD model. 
The similar conclusion is also made in \cite{Bastid:2005ct}, where the directed flow $v_1$ in 1.69A GeV Ru+Ru collisions at SIS/GSI is measured with the cumulant method. It is found that $|v_1\{2\}|$ is less than $|v_1\{4\}|$ which is also contrary to the predictions based on flow fluctuations. 

In spite of the good agreement of $v_2\{4\}$ or $v_2\{6\}$ with the exact $v_2$ in semi-central bins, from Fig.\ref{fig1}, we have also seen that both $v_2\{4\}$ and $v_2\{6\}$ do not agree with the exact $v_2$ in the most central and the very peripheral bins. This means at central and very peripheral collisions, the $v_2$ fluctuations indeed play an important role as indicated in \cite{miller03}. In the peripheral bins the
higher order cumulants give larger $v_2$ than the exact one. In the most central bin, the $v_2\{4\}$ is smaller
and even becomes complex (not shown in Fig.\ref{fig1}) due to the fluctuations, while the $v_2\{6\}$ is slightly
larger than the exact $v_2$. These findings are qualitatively consistent with previous results within a
simplified Monte-Carlo Glauber treatment \cite{miller03}.

In order to estimate how sensitive the cumulant method is to impact parameter fluctuations in a centrality bin,
we also performed the cumulant analysis in enlarged centrality bins. The pink (grey) points in Fig.\ref{fig1}
show the results for the enlarged bins(0-10\%, 5-20\%, 10-30\%, 20-40\%, 30-50\%, 40-60\%, 50-70\%). One can see
that the $v_2$ values from any order cumulants are still in line with the corresponding $v_2$ results from the
original bins although the impact parameter fluctuations in the enlarged bins are larger than those in the
original (narrower) centrality bins.  Thus, the main contribution to the $v_2$ fluctuations in the original
centrality bins should be due to $v_2$ fluctuations at the same impact parameter, e.g. due to the spatial eccentricity
fluctuations from event to event and not due to impact parameter fluctuations.

\begin{figure}[h]
\centering
\includegraphics[totalheight=3.7in]{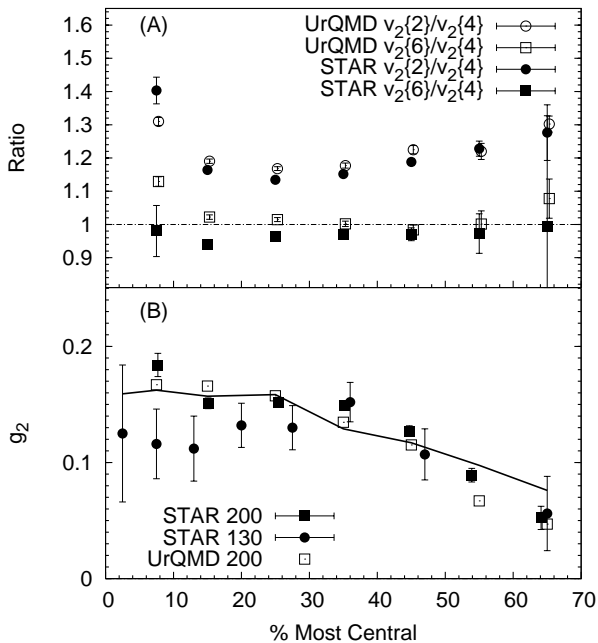}
\caption{(A) The ratios $v_2\{2\}/v_2\{4\}$ and $v_2\{6\}/v_2\{4\}$ from UrQMD are compared to the STAR data
\cite{STAR_flow}. (B) The $g_2$ factors from the UrQMD model are compared to the STAR data, the solid line is the exact $g_2$, i.e. $N\cdot(v_2\{2\}^2-v_2^2)$ (please see the text for details). Note that $g_2$'s from UrQMD have been scaled down by a factor 0.186 } \label{fig2}
\end{figure}

While the total elliptic flow values extracted from the calculation are lower than the experimental results, the
relations between $v_2\{2\}$, $v_2\{4\}$ and $v_2\{6\}$ are similar to the results reported by the STAR
collaboration at RHIC. As shown in Fig.\ref{fig2}(A), open symbols denote the calculation, while full
symbols show the STAR data on the ratios $v_2\{2\}/v_2\{4\}$ and $v_2\{6\}/v_2\{4\}$ for comparison. The good agreement between UrQMD results and the data may indicate that the mechanism which accounts for the differences between $v_2\{2\}$ and $v_2\{4\}$ or $v_2\{6\}$ is the same. 
In Fig.\ref{fig2}(B), we show the $g_2$ factor from the UrQMD model. The $g_2$ factor, is defined in experiments as
\cite{Borghini2002} $g_2=N\cdot(v_2\{2\}^2-v_2\{4\}^2)$, where N is the event multiplicity (for our analysis) or
the number of wounded nucleons (for the STAR data) which should be approximately proportional to the
multiplicity. The $g_2$ is a measure of the non-flow effects and should be independent of the centrality as
originally suggested by \cite{Borghini2002}. However, the STAR \cite{STAR_flow} and SPS \cite{SPS_flow} data
show that with the increase of the impact parameter, the $g_2$ will decrease by about a factor of 3. This decrease
of the observed $g_2$ is consistent with the results based on the eccentricity (or $v_2$) fluctuations
\cite{miller03}, which seems confirm the conjecture in \cite{Borghini2002}, i.e. the $v_2$ fluctuations account for the variance of $g_2$ with centrality. As we can see in Fig.\ref{fig2}(B),
the $g_2$ from the UrQMD model (the blank square) also has similar shape as the data (please note that $g_2$ from UrQMD has been rescaled by a factor 0.186 to compare to the 200AGeV STAR data, since the magnitude of the $v_2$ is too small and N in the $g_2$ defintion is also different).
However, the decrease of $g_2$ in the UrQMD model is not mainly from $v_2$ fluctuations. Since the exact $v_2$ is known in the UrQMD model, the exact two-particle non-flow correlations (or the exact $g_2$), i.e. $N\cdot(v_2\{2\}^2-v_2^2)$, is also known. In Fig.\ref{fig2}(B), the solid line is the exact $g_2$ from the UrQMD model. Apparently, the exact $g_2$ decreases towards peripheral centralities too. Therefore, although the $v_2$ fluctuations indeed make the experimentally measured $g_2$ decrease slightly faster, the non-flow correlations themselves change with the centralities too.

Recently, to overcome the experimental limitations in the $v_2$ measurement with the reaction plane method, the STAR experiment has upgraded its set-up. The Shower Max detector of the Zero Degree Calorimeters(ZDC-SMD) has been added to reconstruct the reaction plane with the sideward deflection (bounce-off) of the spectator neutrons. The non-flow effects are supposed to be minimal, because the spectator neutrons barely participate in the complicated final state rescattering. The STAR preliminary results \cite{GWang_QM05} for the measured $v_2$ with
respect to this reaction plane is denoted as $v_2\{$ZDC-SMD$\}$. The reported $v_2\{$ZDC-SMD$\}$ agrees well with
$v_2\{4\}$ in the mid-central collisions(10-50\%). $v_2\{$ZDC-SMD$\}$ is larger than $v_2\{4\}$ in the most
central bins(0-5\% and 5-10\%) and smaller than the $v_2\{4\}$ in the very peripheral bins(larger than 50\%).
The relation between $v_2\{$ZDC-SMD$\}$ and $v_2\{4\}$ is similar with those between exact $v_2$ and $v_2\{4\}$
from UrQMD. This similarity, on the one hand, confirms that the mechanism which affects the cumulant method is indeed the same as that in UrQMD; on the other hand, it supports that the flow measurement with the ZDC-SMD is not disturbed by
non-flow effects or flow fluctuations. Therefore we want to advocate the ZDC-SMD method for the flow
analysis, because it allows to extract very reliable flow results over the whole centrality.

Let us now turn to the study of the the differential $v_2$. In the cumulant method, the differential $v_2$ in one $p_T$ or rapidity bin is estimated with the cumulants between the particles in this bin and those in one common ``pool''. The average $v_2$ of the particles in the ``pool" should be known from the integral flow analysis. For the following differential $v_2$ analysis, we always use all the particles within $|\eta|<2.5$ as the ``pool". One should also notice that the non-flow correlations which affect the differential flow analysis will be that between the particles in the chosen bin and those in the ``pool".  

\begin{figure}[h]
\centering
\includegraphics[totalheight=2.3in]{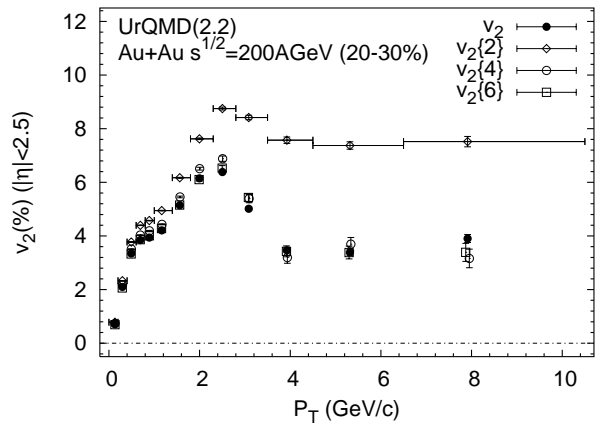}
\caption{$v_2(p_T)$ in the semi-central collisions: results from the cumulant method are compared to the exact
$v_2$} \label{fig3}
\end{figure}

Firstly, let us explore the $p_T$ dependence of $v_2$. Here we use more than $6\cdot 10^5$  semi-central
events (with impact parameters from 6.7 to 8.3~fm corresponding to about 20\% to 30\% of the total cross
section). From the above results on the integral $v_2$, we know that both four and six-particle cumulants
produce almost the exact $v_2$ in this centrality bin, but it is still necessary to see whether the different
cumulant method produce the differential $v_2$ correctly. Especially at large transverse momenta ($p_T$)
non-flow contributions are expected to be large and might influence the results obtained by the cumulant method.
Fig.\ref{fig3} shows the calculations for the $v_2$ of  particles within $|\eta|<2.5$ as a function of $p_T$. At
low $p_T$, the exact $v_2$ increases with the increase of $p_T$ and reaches a maximum at about 2.5~GeV/c, then
drops down with a further increase of $p_T$. In contrast, the $v_2$ from two-particle cumulant $v_2\{2\}$
increases also at low $p_T$, but stays roughly constant at large $p_T$, in addition it is always higher than the
exact $v_2$. The saturation of $v_2\{2\}$ is consistent with STAR's $v_2\{2\}$ results \cite{STAR_highpt}. This
strong deviations point towards substantial contributions from non-flow effects in the two-particle cumulant
method. The higher order cumulants do a much better job in reproducing the exact $v_2$. Here, the difference
between $v_2\{4\}$ and the exact $v_2$ is much smaller especially at large $p_T$. However, $v_2\{4\}$ is still
larger than the exact $v_2$, indicating that even four-particle cumulants are not free from  non-flow
disturbances. When we go to the six-particle cumulant results $v_2\{6\}$, we get good agreement with the exact
$v_2$ in the whole $p_T$ range within the statistical error. This shows that the non-flow effects have been
completely eliminated in $v_2\{6\}$. 

\begin{figure}[h]
\centering
\includegraphics[totalheight=3.7in]{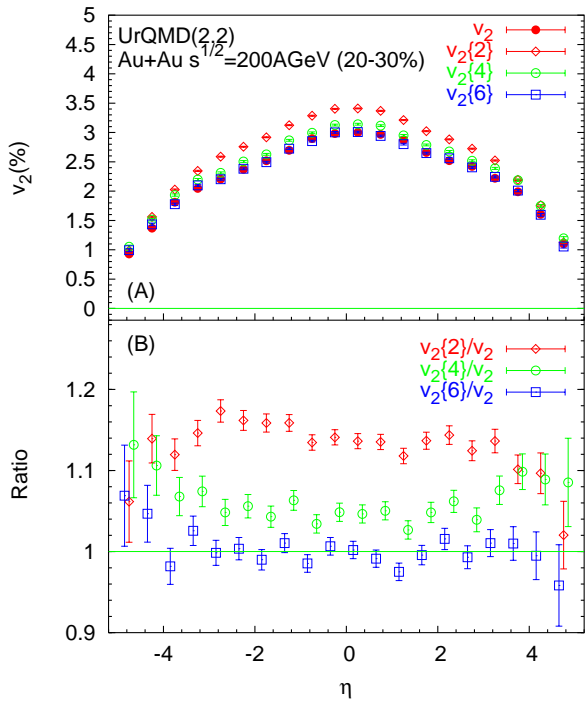}
\caption{(Color online) (A) $v_2(\eta)$ in  semi-central Au+Au collisions at $\sqrt s=200$~AGeV. Results from the cumulant method are compared to the exact $v_2$. (B) Ratios of $v_2\{n\}$ over the exact $v_2$.} \label{fig4}
\end{figure}

Finally, we will study the pseudo-rapidity ($\eta$) dependence of $v_2$ with the cumulant method using the same set of semi-central events as for transverse momentum analysis. It is usually expected that at large $\eta$, the non-flow effects are less important than at mid-rapidity because of the larger rapidity gap between the particles in the rapidity bin and the ``pool" particles. So the difference between $v_2\{2\}$ and $v_2\{4\}$ might be smaller at large $\eta$ compared to midrapidity. Fig.\ref{fig4}(A) shows the results on $v_2(\eta)$ obtained from the different methods. Indeed one observes that at large $\eta$, $v_2\{2\}$, $v_2\{4\}$ and $v_2\{6\}$ are almost similar and they all agree well with the exact $v_2$. This is in line with the STAR results on the $v_2(\eta)$ also indicating agreement between $v_2\{2\}$ and $v_2\{4\}$ at large $\eta$ \cite{STAR_flow}. However, the smaller difference between the $v_2$'s from any-order cumulants at larger rapidity must not be taken as a sign that the non-flow effects are less important at larger rapidities, because the $v_2$ itself decreases towards large rapidity. To demonstrate this, Fig.\ref{fig4}(B) shows the ratios of $v_2\{n\}$ over the exact $v_2$. One observes that the ratios are roughly independent of the rapidity. Therefore, the non-flow effects at forward rapidity might be as important as those at mid-rapidity. 

After the test of the cumulant method, we know that $v_2$ from the two-particle cumulant is heavily affected by the non-flow effects. While the still remaining question is what is the main origin of these non-flow correlations. 
Firstly, in the UrQMD model, there is no correlations due to the HBT effect. The correlations due to the overall transverse momentum conservation are also found to be less important for the $v_2$ measurement \cite{Borghini:2000cm}. The short range direct correlations due to resonance decay, jet production and the last elastic or inelastic interaction before freeze-out can be removed by selecting the particles in the analysis preferentially. That is to say, only one of the final state particles that are {\bf directly} from the above processes is selected in the analysis. 
The integral $v_2\{2\}$ and $v_2\{4\}$ of the selected particles in the semi-central bin (20-30\%) are almost the same as those of the original analysis. Therefore, the short range direct correlations are not the main non-flow correlations at least for this centrality. 
It should be stressed that the jet processes happen at the very beginning of the event and most resonances decay before the freeze-out of the system. Therefore, most of the daughter particles directly from jets or resonances will interact with other particles in the system. And the correlations between these daughter particles could be changed by the final state interactions.  
As has been shown in \cite{Voloshin:2003ud}, the final state rescatterings, or more exactly, the transverse radial expansion can contribute to the non-flow correlations between these daughter particles due to the fact that the particles from the same $NN$ collsions (or jets and resonances) are pushed in the same direction radially in the transverse plane. Ref. \cite{Voloshin:2003ud} predicts that, this kind of non-flow correlations will depend on the centrality (following the development of radial flow). It is indeed observed both in the UrQMD results and in the data that $g_2$ factor increases towards central collisions, which shows the non-flow correlations between the daughter particles of jets and resonances due to transverse radial expansion could be very important or even dominant for semi-central to central collisions. 

In summary, we have applied the cumulant method to analyze the $v_2$ of the Au+Au reactions at $\sqrt{s}=200$~AGeV
within the UrQMD model. On the integral $v_2$ analysis, we reproduce the hierarchy of $v_2\{2\}$, $v_2\{4\}$ and
$v_2\{6\}$ observed by the STAR experiment even if the $v_2$ from UrQMD is only about 60\% of the data. From
the comparisons of the cumulant results to the exact $v_2$, we found that $v_2$ fluctuations affect the results
from the cumulant method in the most central and very peripheral collisions. However, this effect  is almost
negligible over a wide range of the mid-central collisions (about 10-50\% of the total cross section). 

While the
two-particle cumulant results are heavily affected by non-flow effects, non-flow effects can indeed be nearly
eliminated using four and six-particle cumulants. The similarity between STAR data and UrQMD results shows that
the new flow measurements at STAR (using the ZDC-SMD detector) are a good way to obtain $v_2$ values which are not
disturbed by the non-flow effects and $v_2$ fluctuations over the whole centrality range. 

For the differential $v_2$ analysis, the two-particle cumulant method gives a nearly saturated $v_2$ at large $p_T$ in stark contrast to the exact
$v_2$ that drops down rapidly at high $p_T$.  The $v_2$ from four and six-particle cumulants agree well with the
exact $v_2$ especially at large $p_T$. However, there are still some non-flow contributions left in the
four-particle cumulant method so that the $v_2\{4\}$ is always a little (about 4\%) larger than the exact $v_2$.
Finally, we point out that in the present model the non-flow effects at forward rapidity might be as important as those at mid-rapidity.

As for the origin of the non-flow correlations, we find that the short-range direct correlations due to resonance decay, jet production and the last final state interactions before freeze-out are not important for the semi-central to central Au-Au collsions at RHIC. While the indirect correlations between the daughter particles of jets and resonances could be the major non-flow effects for these collisions, especially these correlations could be substantionally enhanced by the transverse radial expansion of the medium \cite{Voloshin:2003ud}.

A final remark. As shown above, the many-particle cumulant method $v_2\{n\ge 4\}$ allows for a good estimation of the exact $v_2$ in the mid-central collisions, thus one may justify other analysis methods by comparing their results with the cumulant method results. For instance, the PHENIX reaction plane method \cite{PHENIX_PRL94} seems also to suffer from non-flow effects because it gives the same $v_2(p_T)$ results as the two-particle cumulant method which is heavily affected by the non-flow effects as discussed above. The STAR ZDC-SMD method seems to give good estimates of the integral flow. But further comparisons with the cumulant method on the differential flow are necessary to fully justify its application in the differential flow analysis. 

\begin{acknowledgments}
We are grateful to the Center for the Scientific Computing (CSC) at Frankfurt for the computing resources. The
authors thank A.~Tang, N.~Xu, G.~Wang, R.~Snellings and Q. Li for helpful and stimulating discussions. This work was supported by GSI
and BMBF. X.Z. thanks the Frankfurt International Graduate School for Science (FIGSS) at the
J.~W.~Goethe-Universit\"at for financial support.
\end{acknowledgments}

\end{document}